\documentclass{article}

\usepackage[final]{tccml_neurips_2020}

\usepackage[utf8]{inputenc} % allow utf-8 input
\usepackage[T1]{fontenc}  % use 8-bit T1 fonts
\usepackage{hyperref}    % hyperlinks
\usepackage{url}      % simple URL typesetting
\usepackage{booktabs}    % professional-quality tables
\usepackage{amsfonts}    % blackboard math symbols
\usepackage{nicefrac}    % compact symbols for 1/2, etc.
\usepackage{microtype}   % microtypography
\usepackage{graphicx}
\usepackage{siunitx}

\setcitestyle{authoryear}

\title{Machine Learning Climate Model Dynamics: Offline versus Online Performance}

\author{%
  Noah D. Brenowitz\thanks{Corresponding author}, Brian Henn, Jeremy McGibbon, Spencer K. Clark, Anna Kwa\\{\bf W. Andre Perkins, Oliver Watt-Meyer, Christopher  S. Bretherton} \\
  Vulcan, Inc.\\
  Seattle, WA \\
  \texttt{\{noahb, brianh, spencerc, annak, jeremym, andrep, oliwm, chrisbr\}@vulcan.com}
}

\begin{document}

\maketitle

\begin{abstract}
  Climate models are complicated software systems that approximate atmospheric and oceanic fluid mechanics at a coarse spatial resolution.
  Typical climate forecasts only explicitly resolve processes larger than \SI{100}{km} and approximate any process occurring below this scale (e.g. thunderstorms) using so-called parametrizations.
  Machine learning could improve upon the accuracy of some traditional physical parametrizations by learning from so-called global cloud-resolving models.
  We compare the performance of two machine learning models, random forests (RF) and neural networks (NNs), at parametrizing the aggregate effect of moist physics in a \SI{3}{\km} resolution global simulation with an atmospheric model.
  The NN outperforms the RF when evaluated offline on a testing dataset.
  However, when the ML models are coupled to an atmospheric model run at \SI{200}{\km} resolution, the NN-assisted simulation crashes with 7 days, while the RF-assisted simulations remain stable.
  Both runs produce more accurate weather forecasts than a baseline configuration, but globally averaged climate variables drift over longer timescales.
\end{abstract}

\section{Introduction}

Machine learning has the potential to improve the accuracy of climate models but should take advantage of our existing physical knowledge.
Climate and weather models represent the motions of the Earth's atmosphere as a system of discretized ordinary differential equations (ODEs).
Some of these terms are well-known from first principles such as the Navier-Stokes equations and radiative transfer.
However, climate models typically have a horizontal grid-size of \SI{100}{\km} \citep{IPCC} and cannot resolve the dominant physical scales of some atmospheric processes, especially turbulence, cumulus convection, and cloud-radiation interactions; these must be parametrized. We call those processes that involve the formation of clouds and precipitation the `moist physics'.
New simulations called global cloud resolving models (GCRMs), which explicitly resolve key aspects of global cloud fields associated with precipitating cumulus convection by using horizontal grid scales of less than 5~km \citep{Satoh2019-yj}, are currently too expensive for long climate simulations but contain a wealth of information about how clouds interact with the large-scale---a valuable training dataset for inexpensive machine learning (ML) parametrizations.

Sub-grid-scale parametrization is a function approximation problem.
Let the ODEs describing a climate model can be divided into two components as follows
\begin{equation}
  \frac{d \mathbf{x}_i}{dt} = \mathbf{g}_i(\mathbf{x}, t) + \mathbf{f}(\mathbf{x}_i; \theta).\label{eq:1}
\end{equation}
The known physics $\mathbf{g}$ account for large-scale atmospheric fluid mechanics as well as some set of known physics, and are a function of the non-local state $\mathbf{x}$.
The vector $\mathbf{x}$ represents the state of the global atmosphere (e.g. temperature, humidity, and wind).
Because the atmosphere mixes rapidly in the vertical direction, parametrized physics $\mathbf{f}$ are typically assumed to depend only on some ML parameters $\theta$ and the \emph{atmospheric column} overlying a single horizontal grid cell $i$.
We denote this horizontally local state with $\mathbf{x}_i \in \mathbb{R}^{m}$ where $m$ is the number of vertical grid points times the number of 3D fields input to the parametrization.
In this work, $f$ will include sources and sinks of humidity and heat due to moist physical processes.

Moist physics are conventionally handled by a suite of human-devised parametrizations encorporating physical constraints and empirical knowledge.
These typically assume an analytical sub-grid-scale cloud model in statistical equilibrium with the large-scale environment \citep{Arakawa2004-io}, but
recent work proposes parameterizing moist physics using machine learning models trained from either higher-fidelity simulations \citep{Rasp2018-ff, Brenowitz2018-td,Brenowitz2019-qs,Yuval2020-ks,Krasnopolsky2010-nn} or reanalysis data \citep{McGibbon2019-ah}.

These ML schemes are trained offline, as a supervised learning problem, where the inputs $\mathbf{x}_i$ and outputs $\mathbf{f}$ are taken from a pre-computed dataset.
Because this training fails to account for feedbacks between $\mathbf{g}_i$ and $\mathbf{f}$, offline accuracy does not translate immediately into online accuracy when the ML is coupled to the fluid dynamics solver and used to simulate the weather or climate.
For example, offline training will often yield a numerically unstable scheme that causes an online simulation to crash within a matter of days \citep{Brenowitz2019-qs,Brenowitz2020-po}.
The reasons behind this are starting to emerge.
\citet{Brenowitz2020-po} demonstrated with formal stability analysis that this instability is related to the linearized behavior of a neural network (NN) when coupled to idealized wave dynamics.
Also, feedback loops in coupled simulations can easily generate out-of-sample inputs unlike any in the training sample.
On the other hand, random forests (RFs) are more stable online, likely because they can only predict averages of observed samples \citep{OGorman2018-hn, Yuval2020-ks}.

Prior work has focused on idealized aqua-planet configurations over a global ocean with fixed surface temperature. While this is a useful prototyping configuration, ML parametrizations need to be accurate with realistic geography and topography to be used in real-world forecast models. Therefore, this manuscript has two goals: 1) demonstrate the feasibility of ML parametrization on a more realistic atmospheric model and 2) compare the offline and online performance of RFs and NNs for this problem.

We first introduce the atmospheric model we are trying to improve and which we use to generate the training data in Section \ref{sec:methods}.
Then, we explain the random forest and NN training processes and ML formulation. 
We share online and offline performance results in Section \ref{sec:results} and conclude in Section \ref{sec:conclusion}.

\section{Methods \label{sec:methods}}

\subsection{The FV3GFS Atmospheric model}

We use the FV3GFS \citep{Harris2020-gm} atmospheric model to generate both fine-resolution training data and evaluate the accuracy at a coarser resolution.
FV3GFS solves the three-dimensional Euler equations over a spherical geometry discretized on a cubed-sphere grid \citep{Putman2007-mk}.

To generate our training data, we run the FV3GFS with an approximate horizontal grid of \SI{3}{\km}.
At this resolution, FV3GFS can resolve the dominant motions due to deep convection \cite{Satoh2019-yj}, so this model is run with only the microphysics scheme, radiative transfer, shallow, convection, and a boundary layer turbulence scheme \citep{Harris2020-gm}.
Since this model has its own set of biases, we include an additional Newtonian relaxation term to the temperature, pressure, and wind fields which nudges this simulation towards a an observationally-derived gridded data product called a reanalysis with a time scale of 1 day.
We perform a 40 day run with this configuration at a NOAA super-computing facility and save the full 3D state of the atmosphere and process-wise tendency information, horizontally block-averaged to \SI{200}{km} resolution, every 15 minutes.

The baseline model we are hoping to improve is based on the FV3GFS model run at a \SI{200}{\km} resolution.
This scale is typical for climate models, but cannot resolve many moist atmospheric processes.
We, therefore, assume that the known physics $\mathbf{g}$ only includes advection together with the standard parametrizations of clear-sky radiation and turbulence, a configuration we call Clouds-off. 
The ML will predict the remaining processes $\mathbf{f}$.
We compare ML-assisted runs with a baseline configuration (All-physics) with the standard human-designed moist physics parametrizations.

\subsection{Machine learning models}

For simplicity, the ML models will only predict sources of temperature and humidity, letting the clouds-off physics handle frictional processes.
We compute the terms-to-be-parametrized $\mathbf{f}$ as budget residual of (\ref{eq:1}), so that $\mathbf{f}_i = d \mathbf{x_i}/dt - \mathbf{g_i} $.
The total tendency $d \mathbf{x_i}/dt$ is the sum of all the physical-process tendencies in the fine-resolution data and the convergence of vertical temperature and humidity fluxes.
If initialized with the coarsened state of the fine-resolution model, the coarse-resolution model will develop strong transients damaging the estimated known physics $\mathbf{g}_i$ \citep{Lynch2008-zu}.
To generate smooth estimates of $\mathbf{g}_i$ and $\mathbf{x}_i$, we  nudge a coarse clouds-off simulation towards the fine-resolution data with a 3-hour nudging timescale for temperature, humidity, pressure, and the winds.
The inputs features $\mathbf{x}_i$ include the vertical profiles of temperature and humidity along with the cosine of the solar zenith angle, the surface elevation, and the land-sea mask.
The nudging time-scale is a regularization parameter; longer time-scales will give smoother, but more biased estimates, of $\mathbf{x}_i$ and $\mathbf{g}_i$.

We compare the online and offline performance of RFs and NNs on this problem.
The training data consist of 130 randomly drawn snapshots from (August 5 through 31) and testing data are 72 snapshots from (September 1 through 7).
Each snapshot contains \num{13824} spatial samples.
Both the RF and NN are trained to minimize the mean-squared-error scaled by the inverse standard deviation of each output feature.
A RF is fit with a maximum depth of 13 and 13 ensemble members (one per batch of 10 timesteps).
A two layer NN, with 128 nodes per layer and ReLU activation, is fit with the Adam optimizer (learning rate 0.001) with inputs normalized by the standard deviation and mean computed over a single batch.
A mini-batch size of 512 is used for NN training and 8 passes through the training data (epochs) are completed.

\section{Results \label{sec:results}}

\begin{figure}
  \centering
  \includegraphics[width=.7\textwidth]{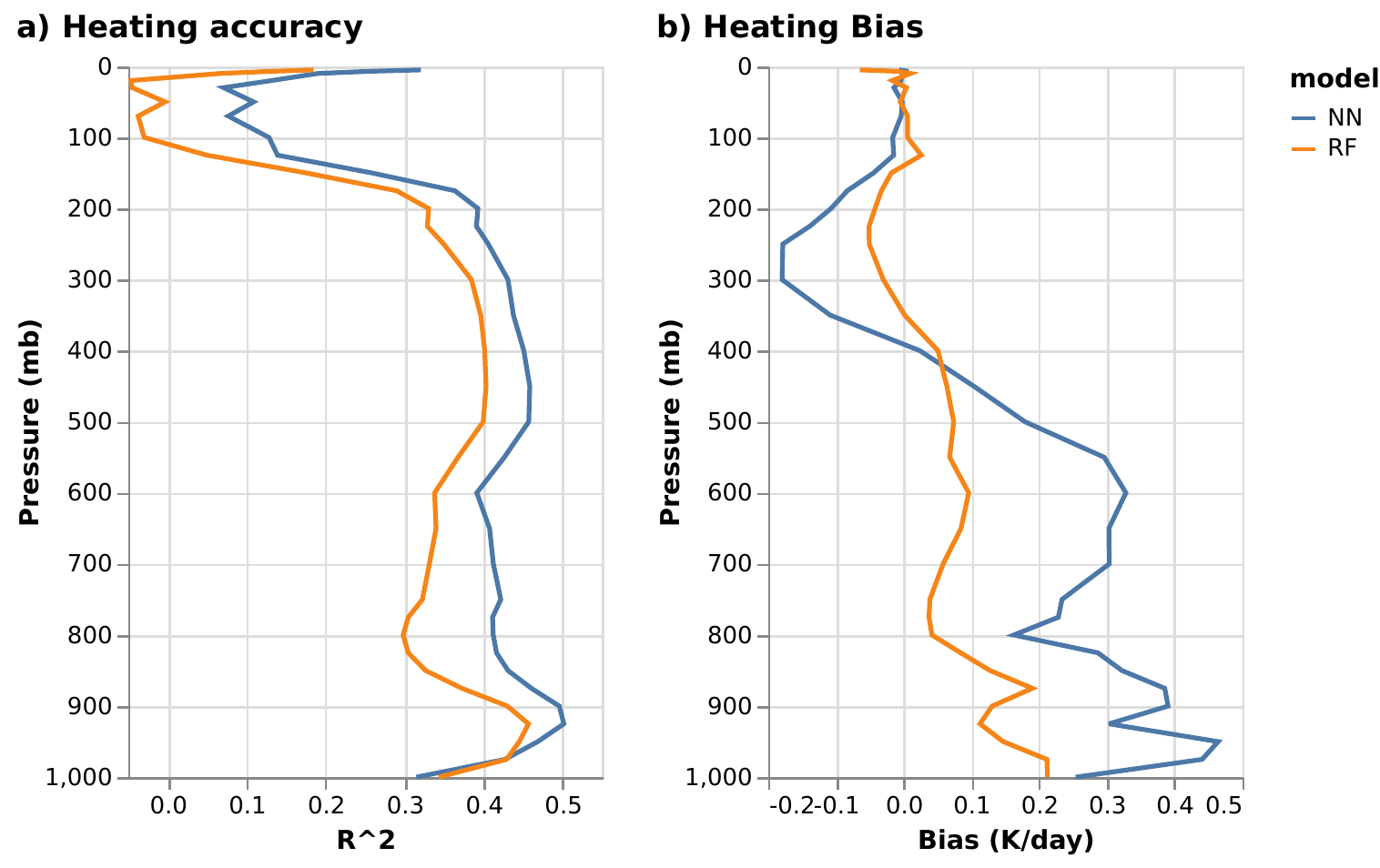}
  \caption{Offline accuracy of ML-predicted temperature source (i.e. heating) for the NN and RF. (a) area-weighted $R^2$ scores and (b) area-weighted averages computed over the testing times. The predictions and truth are interpolated to fixed pressure levels before computing the metrics.\label{fig:offline}}
\end{figure}

The RF and NN have comparable accuracy offline on the testing data (cf. Figure \ref{fig:offline}).
The RF has a worse coefficient of determination ($R^2$) than the NN does at all pressure levels of the output.
On the other hand, the global and time average of the NN is more biased than the RF, likely because this problem features highly non-Gaussian outliers that distort the MSE-based loss function of the NN.
Both the RF and NN predict too much global average heating in for pressure levels between \SIrange{1000}{400}{mb}.

\begin{figure}
  \centering
  \includegraphics[width=\textwidth]{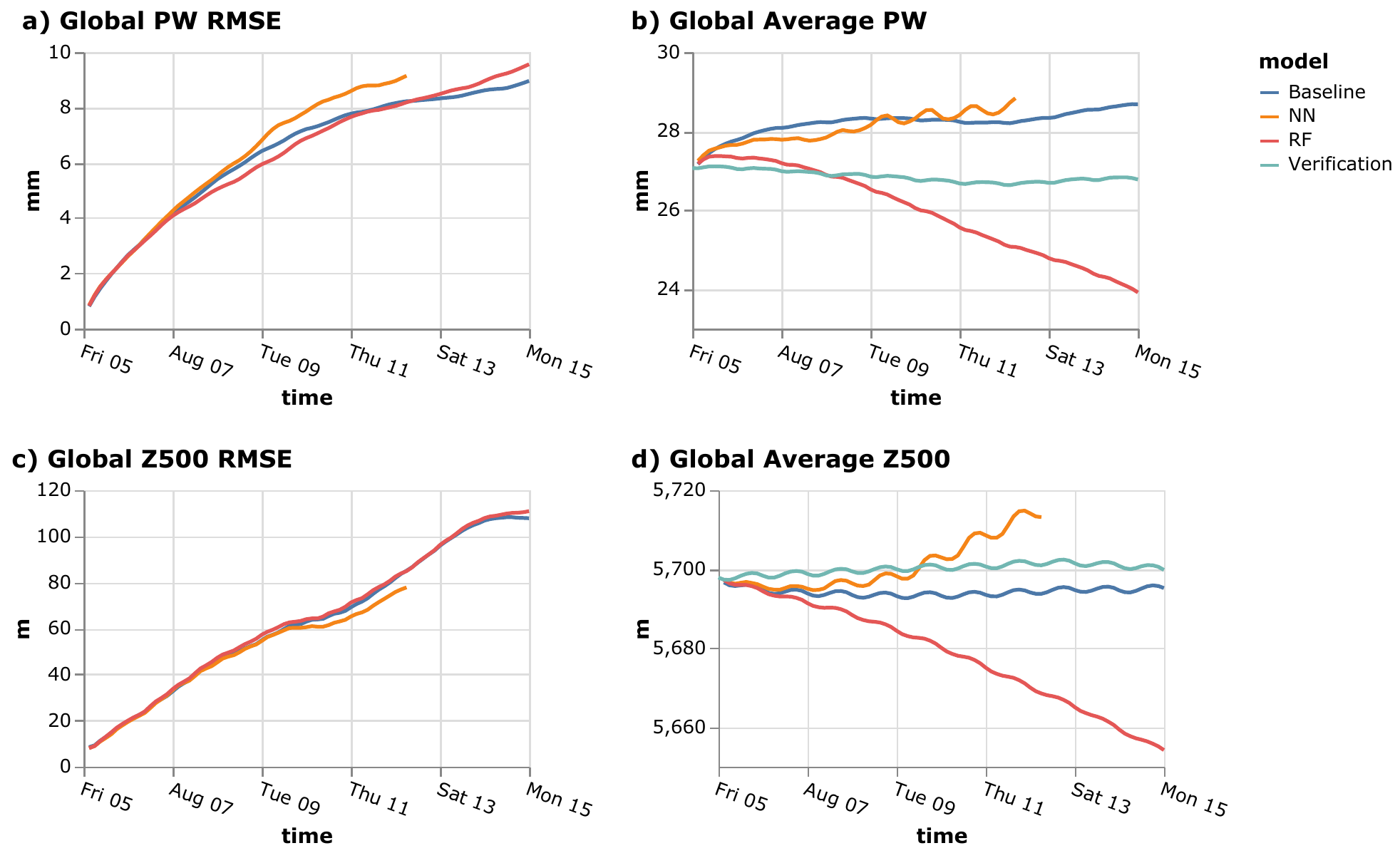}
  \caption{Online skill of the NN and RF. Compares the forecast accuracy in terms of RMSE for precipitable water (a) and \SI{500}{mb} geo-potential height (c). The respective global averages are shown in (b) and (d).}
  \label{fig:online}
\end{figure}

To test the online accuracy, we couple the NN and RF to the \SI{200}{\km} atmospheric model run with clouds-off known physics, and perform 10-day hindcast simulations initialized at 0 UTC on August 5, 2016 and compare them to an all-physics simulation (cf. Figure \ref{fig:online}).
Even though the NN outperformed the RF offline, the NN simulation crashes after around 7 days, while the RF simulation successfully completes the 10-day simulation.
This confirms findings that random forests are often stable for this problem \citep{OGorman2018-hn} while neural network are not \citep{Brenowitz2020-po}.

We evaluate the predictive skill for each ML method at each forecast lead time using the root-mean squared error (RMSE) and global average of precipitable water (PW) and \SI{500}{mb} geopotential height (Z500) compared to the verification high-resolution training data across all grid columns around the globe.
Overall, the global average of Z500 for the NN remains closer to the verification until it crashes, and it has the highest forecast skill of the three simulations at predicting Z500.
The RF has the lowest RMSE and smallest bias for PW for a few days, but eventually starts to dry out.
Overall, the baseline model is more robust and has less systematic drift in PW and Z500, but the ML-assisted runs have better skill (lower RMSE) for forecasts for up to 5 days.

\section{Conclusion \label{sec:conclusion}}

We have compared how well a random forest and a simple neural network perform when they replace the human-designed moist-physical parametrizations of a coarse resolution model.
Unlike past studies, we have trained the RF and NN with the exact same training data, a global cloud-resolving simulation with an approximate resolution of 3km.
To our knowledge, this is the first such clean comparison in the literature.\footnote{Since the initial submission of pre-print, \citet{Yuval2020-tc} have also performed a similar comparison.}
Moreover, this simulation is significantly more complex that the idealized datasets used in past work because it includes a realistic land surface and topography.

The NN is more accurate offline than the RF, but is not numerically stable online (i.e. when coupled to known physics).
Because the training process does not account for feedbacks between the ML and the known physics, online simulations can quickly produce samples unlike any seen in the training data.
Because RFs predict outputs within the convex hull of their training data, they are likely more robust than NNs when forced to extrapolate to such new samples.
Future work should focus on finding an offline input-output prediction problem that translates to good online performance.

The global average of the NN predictions are also more biased offline, possibly because its training procedure is less robust to extreme rainfall events in the training data.
This offline bias could possibly be addressed using robust loss functions (e.g. Huber loss or mean absolute error).

\section*{Broader Impact}

If successfully expanded upon, this work promises to improve the physical models we use to forecast weather and climate with machine learning.
In particular, our goal is to improve accuracy of precipitation forecasts with these ML moist physics parametrizations.
Such forecasts, with quantified uncertainty, of precipitation trends and extremes in a changing climate will allow policymakers and the general public to make better-informed decisions about climate impacts on many aspects of society and the natural world.

\begin{ack}
The authors acknowledge the support of Vulcan, Inc. for funding this project.
We thank Lucas Harris for support with the FV3GFS model and the Geophysical Fluid Dynamics Laboratory (NOAA) for super-computing resources.
\end{ack}

\bibliography{references}

\end{document}